\documentclass[12pt]{iopart}
\usepackage{graphicx}
\usepackage{amssymb}
\usepackage{bm} 
\usepackage{braket} 

\expandafter\let\csname equation*\endcsname\relax
\expandafter\let\csname endequation*\endcsname\relax 
\usepackage{mathtools} 

\begin{document}

\title{Disorder effects on the coupling strength of coupled photonic crystal slab cavities}

\author{J.P. Vasco and V. Savona}

\address{Institute of Physics, \'Ecole Polytechnique F\'ed\'erale de Lausanne (EPFL), CH-1015 Lausanne, Switzerland}
\ead{juan.vasco@epfl.ch and vincenzo.savona@epfl.ch}
\vspace{10pt}

\begin{abstract}
We study the effects of disorder on the coupling strength of coupled photonic crystal slab cavities by considering fully-3D electromagnetic calculations. Specifically, we investigate two coupled $L3$ cavities at 30$^\circ$ and 60$^\circ$ configurations, where the coupling strength $J$ (or photon hopping) is extracted from the simulations in the presence of disorder. We found that the relative fluctuations of the photon hopping are more sensitive to disorder effects than the corresponding fluctuations in the eigenfrequencies of the coupled cavities. Furthermore, for the typical range of disorder in state-of-the-art devices, the $J$ fluctuations are found to increase linearly as a function of the disorder amplitude. This allows to set upper bounds to the amplitude of fabrication imperfections, for which the coupling predicted by design can still be expected, on average, in a fabricated device. 
\end{abstract}

%
\vspace{2pc}
\submitto{\NJP}
%
%
%

\section{Introduction}

Photonic crystals (PhCs) have been the focus of intense investigations over the last three decades, thanks to the unique possibility they offer to engineer the density of states of the electromagnetic field and, consequently, the flow of light \cite{yablonovitch,john}. The maturity subsequently achieved in semiconductor nano-fabrication technology has led to a plethora of PhC-based device concepts that may have large impact in future integrated photonic platforms \cite{noda1,vuckovicrev,notomirev,gallirev,lodahl}. In analogy with electron states in crystals, the electromagnetic modes in a PhC form bands. When in presence of a forbidden bandgap, defects in PhCs can be engineered to give rise to confined light modes, typically in 0-D (cavities) or 1-D (waveguides) \cite{joannopoulosBook}. In particular, two-dimensional PhCs made of a triangular lattice of holes on a dielectric slab, have emerged as a paradigm for various functionalities, as they enable the realization of cavities and waveguides with various geometries and figures of merit, to be used as building blocks of more complex photonic devices. Important recent examples are large-scale integration of all-optical memories \cite{notomi1}, all-optical PhC logic gates and switching \cite{liu,imamoglu,notomi2}, strong coupling between distant PhC nanocavities \cite{noda2}, optical sensing \cite{salemink1,salemink2} and wide-band slow light coupled-cavity waveguides \cite{momchil1,badolato}. Moreover, recent advances in PhC optimization have shown that figures of merit -- such as Q-factors or broadband slow light -- can be dramatically improved with simple local changes to the nominal PhC structure \cite{badolato,momchilsr,galli,momchilgan,ulagalandha,momchilwv,momchill43,mohamed}.

Even with the most advanced nano-fabrication resources, disorder in the form of small deviations from the nominal geometry is always present, to some extent, in fabricated PhC structures. The effects of structural disorder on the properties of PhC structures have been extensively studied in the case of cavities \cite{ramunno,momchil2,vasco1} and waveguides \cite{dario,hughes,momchil3,vasco2}, as well as in coupled cavity systems \cite{dignam,savona1,vuckovic}. Most of these studies have been focusing on the effect of disorder on the resonant frequencies of the electromagnetic modes and on their quality factors. The question of how disorder affects the coupling induced by spatial proximity in the case of two resonant PhC structures, remains yet to be addressed. More specifically, if the coupled cavity system is described by the simplest coupled-mode model -- with resonant frequencies $\omega_{1,2}$ respectively for the two cavities, and a photon hopping rate $J$ -- we are interested in the fluctuations of $J$ for a given statistical ensemble of disorder realizations.

In this work we investigate the disorder effects on the coupling strength of coupled PhC cavities. By means of the guided mode expansion method (GME) and a two-coupled-mode model, we compute the coupling strength $J$ of two coupled $L3$ PhC cavities in the presence of disorder, i.e., in the regime where the two cavities have different resonant frequencies. We find that the relative fluctuations of $J$ with respect to their mean values, $\delta J/\braket{J}$, are more sensitive to disorder than the relative fluctuations of the normal mode frequencies, $\delta \Omega/\braket{\Omega}$, of the coupled cavity system. We finally establish the scaling law of $\delta J/\braket{J}$ as a function of the disorder parameter and set an upper bound to the amplitude of random imperfections in the system, in order for the nominal coupling to be realized on average, i.e. $\delta J/\braket{J}\ll1$. The present results are of particular relevance for the design of state-of-the-art integrated PhC devices and for the rapidly growing research on quantum many-body phenomena in photonic arrays \cite{hartmann1,savona2,cristiano,hugo,vasco3,angelakis}.

The paper is organized as follows. In section~\ref{system-methods} we discuss the resonant modes of the ideal (i.e. non disordered) system of two coupled PhC cavities, and explain the procedure to compute the coupling strength $J$, by combining GME with a two-coupled-mode model in the presence of disorder. In section~\ref{results}, we present results for the coupled cavity system in presence of disorder and discuss the scaling law for the relative fluctuations $\delta J/\braket{J}$ and $\delta \Omega/\braket{\Omega}$. Finally, we draw conclusions in section~\ref{conclusions}.

\begin{figure}[t]
  \begin{center}
    \includegraphics[width=0.99\textwidth]{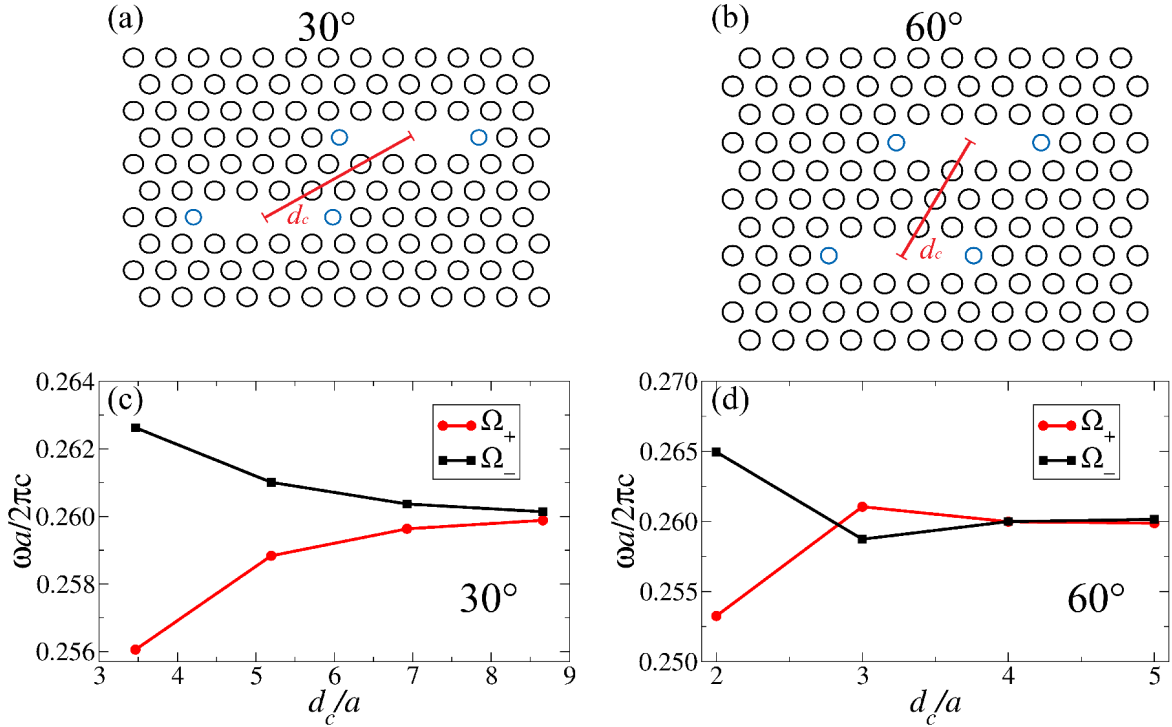}
  \end{center}
  \caption{(a) Schematic representation of the photonic molecule formed by two-coupled $L3$ cavities where the line connecting the cavity centers makes and angle of 30 degrees with the horizontal axis. The cavity-cavity separation, measured from the cavity centers, is denoted as $d_c$. The lateral holes of the cavities (in blue) are displaced outwards by $0.15a$ and their radii reduced to 80\% of the original size. The resulting quality factor for the fundamental mode of each cavity is $Q=106500$. (b) Same as (a) for the case in which the line connecting the cavity centers makes and angle of 60 degrees with the horizontal axis. (c) Frequencies of the normal modes of the 30$^\circ$ molecule, where the even $\Omega_+$ and odd $\Omega_-$ modes are represented by the red and black curves, respectively. (d) Same as (c) for the 60$^\circ$ configuration.\label{fig1}}
\end{figure}

\section{System and methods}\label{system-methods}

We consider the system of two coupled $L3$ PhC cavities, as schematically shown in Fig.~\ref{fig1}, which we will refer to as a photonic molecule. Two specific configurations are investigated: the one in which the line connecting the centers of the cavities makes an angle of 30 degrees with respect to the horizontal axis, Fig.~\ref{fig1}(a), and the one in which this angle is 60 degrees, Fig.~\ref{fig1}(b). The cavity-cavity distance, measured from the cavity centers, is denoted by $d_c$. We consider parameters that are typical for silicon PhCs, i.e., refractive index $n=3.46$, hole radii $r=0.25a$ and slab thickness $d=0.55a$ where $a$ is the lattice parameter of the underlying hexagonal PhC lattice. With this choice, if the lattice parameter is set to $a=400~\mbox{nm}$, the resonant wavelength of the L3 cavity lies around $\lambda=1.55~\mbox{$\mu$m}$. We employ a partially optimized design of the $L3$ cavities where the lateral holes [blue holes in Figs.~\ref{fig1}(a) and \ref{fig1}(b)] are displaced outwards along the horizontal direction by $s=0.15a$ and their radii decreased to $r^\prime=0.8r$. In this way, the nominal (unloaded) quality factor of a single L3 mode is $Q=106500$. The normal modes of the PhC molecule are computed using the GME approach, in which the Bloch modes of the PhC are expanded on the basis of guided modes of the effective homogeneous semiconductor slab \cite{gme}. For the simulations, periodic boundary conditions are assumed, with a rectangular supercell of lateral dimensions $20a\times7\sqrt{3}a$ and $17a\times8\sqrt{3}a$ for the 30$^\circ$ and 60$^\circ$ cases, respectively. For the expansion onto guided modes, a momentum cutoff $a|\mathbf{G}|_{\rm max}=20$ is used, and only the first TE guided mode of the slab was included in the GME expansion. We have carefully verified the convergence of the results with respect to these simulation parameters. The frequencies of the two normal modes originating from the coupling between the two fundamental $L3$ cavity modes are shown for the ideal 30$^\circ$ configuration in figure~\ref{fig1}(c), and for the 60$^\circ$ one in Fig.~\ref{fig1}(d). In the absence of disorder, these modes are respectively even and odd with respect to the inversion symmetry around the middle point of the structure. A coupling arises from the overlap between cavity modes, giving rise to a normal mode splitting $\Delta$ which depends on the cavity-cavity separation $d_c$. Interestingly, spatial oscillations in the tails of the overlapping cavity modes can give rise to a non-monotonic dependence of $\Delta$ as a function of $d_c$. This phenomenon is evidenced in Fig.~\ref{fig1}(d) for the 60$^\circ$ case. Eventually, for large cavity-cavity separation, both normal modes tend to the single frequency value of the fundamental $L3$ cavity mode \cite{vasco4,chalcraft}. 

\begin{figure}[t]
  \begin{center}
    \includegraphics[width=0.99\textwidth]{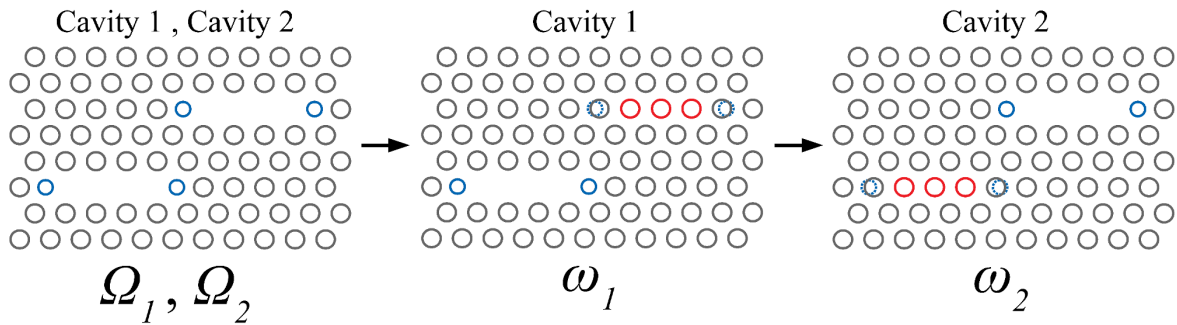}
  \end{center}
  \caption{Schematic representation of the algorithm employed to compute the normal mode and cavity frequencies in the coupled system. First, $\Omega_1$ and $\Omega_2$ are computed for a specific disorder realization. Second, for exactly the same disorder realization, cavity 2 is filled (red holes) with non-disordered holes, and the lateral ones are brought back to their regular positions and sizes (the original holes are shown in dotted-blue), to compute $\omega_1$. Finally, cavity 1 is filled and their lateral holes regularized (in the same way as it was done in cavity 2), to compute $\omega_2$.\label{fig2}}
\end{figure}

Fabrication disorder in the PhC structure is modeled in terms of independent and non-correlated random fluctuations of all hole positions and radii within the supercell \cite{momchil2,nishan}. The probability distribution of the random fluctuations is chosen to be Gaussian with the corresponding standard deviation $\sigma$ taken as our disorder parameter, in line with previous experimental and theoretical characterizations of intrinsically disordered PhCs \cite{galli,dario,thomas,garcia,vasco2}. When disorder is present, extrapolating the value of the coupling strength $J$ from the GME simulation is not straightforward. Indeed, disorder produces both a fluctuation in the resonant frequencies of the two uncoupled cavities and a fluctuation in the coupling strength $J$. In order to extract the value of $J$ from the simulation we adopt the following procedure based on a coupled mode theory. The two normal modes of the molecule are modeled as the eigenmodes of the $2\times2$ matrix
\begin{equation}\label{2x2m}
 M = \begin{pmatrix}
  \omega_1 & J \\
  J & \omega_2
 \end{pmatrix},
\end{equation}
where $\omega_1$ and $\omega_2$ are the (yet unknown) resonant frequencies of the two uncoupled cavities. The idea consists in determining -- for each given realization of disorder -- the value of $J$ for which the eigenvalues of the matrix $M$ are closest to the GME-computed normal-mode frequencies, which we denote by $\Omega_1$ and $\Omega_2$ (assuming $\Omega_1<\Omega_2$). To this purpose, we need to compute the values of the uncoupled resonant frequencies $\omega_1$ and $\omega_2$, for the very same disorder realization. The values $\Omega_1$, $\Omega_2$, $\omega_1$, and $\omega_2$ are all obtained from GME calculations, following the steps sketched in Fig. \ref{fig2}. For a given disorder realization, we first use GME to simulate the two coupled cavities, thus obtaining the eigenfrequencies $\Omega_1$ and $\Omega_2$. Then, for the same disorder realization, we restore the missing holes of cavity 2 (red circles) and bring back the nearest side holes to their regular positions and sizes. This corresponds to effectively removing cavity 2 from the given disordered realization. Simulating this new configuration with GME provides the uncoupled frequency $\omega_1$ of cavity 1. Finally, by following the same steps as for cavity 1, we compute the frequency $\omega_2$. One important question concerns the sensitivity of $\omega_1$ and $\omega_2$ to disorder in the added holes. We would expect that disorder on the three restored holes of cavity 1 (2) should have minor influence on the GME-computed frequency of cavity 2 (1), as the restored holes reside far from the position where the resonant cavity mode is confined. We have checked this assumption by actually considering several disorder realizations for the restored holes. It turns out that disorder on the three restored holes produces fluctuations in the computed resonant frequencies which are always negligible with respect to the main fluctuations induced by disorder on the whole supercell. We can therefore safely avoid to average over several realizations of disorder on the restored holes and simply introduce them in their nominal positions and radii. In all results shown below, we have considered 500 independent disorder realizations and we have checked that the first and second moments of the computed statistical distributions are well converged for this number of realizations. Once the frequencies $\Omega_1$, $\Omega_2$, $\omega_1$, and $\omega_2$ are know for a given disorder realization, the coupling strength $J$ is determined by fitting the eigenfrequencies of the coupled-mode model (\ref{2x2m}) to these values. This is achieved by minimizing the function
\begin{equation}\label{costf}
F(J)=\sqrt{\left[\lambda_1(J)-\Omega_1\right]^2+\left[\lambda_2(J)-\Omega_2\right]^2},
\end{equation}
where $\left\{\lambda_1(J),\lambda_2(J)\right\}$ are the eigenvalues of the matrix $M$. It may be argued that the coupling in the photonic molecule under consideration is not adequately modeled by a coupled-mode theory with only one mode per cavity, and that the effect of the presence of other resonant modes plays a role. We ruled out this possibility by generalizing the procedure described above, to include higher cavity modes in a coupled-mode theory with a larger number of modes. The generalized approach produced values of $J$ that are essentially unchanged with respect to the simple two-mode approach, thus confirming the validity of this latter.

\begin{figure}[t]
  \begin{center}
    \includegraphics[width=0.99\textwidth]{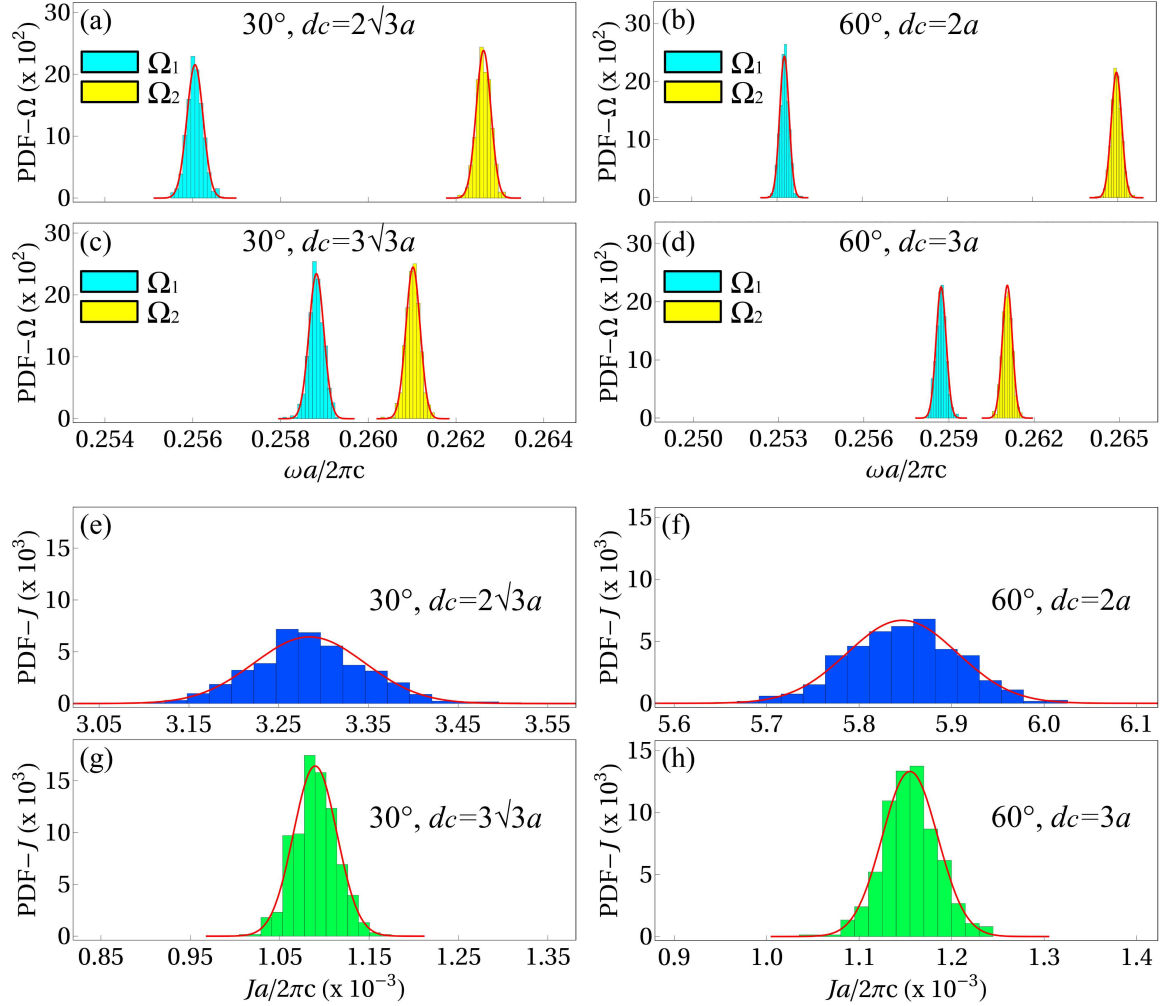}
  \end{center}
  \caption{(a) Probability density function (PDF) of the normal modes $\Omega_1$ (cyan) and $\Omega_2$ (yellow) for $d_c=2\sqrt{3}a$ in the 30$^\circ$ configuration. (b) PDF of the normal modes $\Omega_1$ (cyan) and $\Omega_2$ (yellow) for $d_c=2a$ in the 60$^\circ$ configuration. (c) Same as (a) for $d_c=3\sqrt{3}a$. (d) Same as (b) for $d_c=3a$. (e) Probability density function (PDF) of the coupling strength $J$ for the 30$^\circ$ configuration at $d_c=2\sqrt{3}a$. (f) PDF of $J$ for the 60$^\circ$ configuration at $d_c=2a$. (g) Same as (e) for $d_c=3\sqrt{3}a$. (h) Same as (f) for $d_c=3a$. In all plots, the red curves are the corresponding Gaussian distributions computed with the averaged values, $\braket{\Omega}$ and $\braket{J}$, and standard deviations, $\delta\Omega$ and $\delta J$. 500 independent realizations of the disordered system where employed with $\sigma=0.005a$.\label{fig3}}
\end{figure}

\begin{figure}[t]
  \begin{center}
    \includegraphics[width=0.99\textwidth]{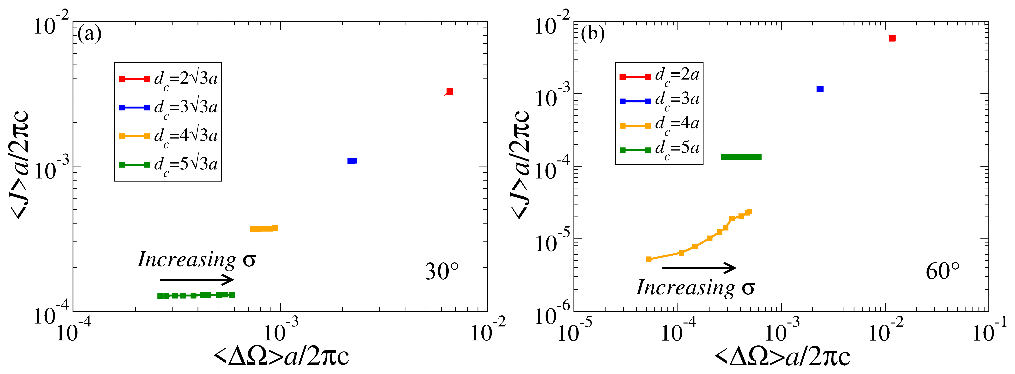}
  \end{center}
  \caption{(a) Averaged coupling strength $\braket{J}$ as a function of the averaged normal mode splitting $\braket{\Delta\Omega}$ for the 30$^\circ$ configuration at different cavity-cavity separations. The corresponding set of points (with the same color) are associated to the different values of $\sigma$, which increases from left to right. (b) Same as (a) for the 60$^\circ$ configuration. Each data point was computed by averaging over 500 independent statistical realizations of the disordered system.\label{fig4}}
\end{figure}

\section{Results}\label{results}

We show in Fig.~\ref{fig3}(a)-\ref{fig3}(d) the probability density function (PDF) of the normal mode frequencies in the 30$^\circ$ and 60$^\circ$ configurations, as computed with GME for a disorder amplitude $\sigma=0.005a$ and two different values of $d_c$; the red lines are fits to the Gaussian distribution, with average $\braket{\Omega}$ and standard deviations $\delta \Omega$. The corresponding PDFs of the coupling strength $J$, and the associated Gaussian fits, are displayed in Figs.~\ref{fig3}(e)-\ref{fig3}(h) for the 30$^\circ$ and 60$^\circ$ molecules.

The link between the averaged coupling strength $\braket{J}$ and averaged normal mode splitting $\braket{\Delta\Omega}$ is shown in Figs.~\ref{fig4}(a) and \ref{fig4}(b), for the 30$^\circ$ and 60$^\circ$ molecules respectively, in log-log scale. We have considered values of the disorder amplitude ranging between $\sigma=0.001$ and $\sigma=0.01$, consistent with the typical accuracy achieved in state-of-the-art semiconductor fabrication \cite{galli,garcia}. The coupling strength $\braket{J}$ increases with increasing $\braket{\Delta\Omega}$, as expected. For a given value of $\braket{J}$, disorder induces an overall increase in $\braket{\Delta\Omega}$, which is more pronounced for small values of $\braket{J}$. This increase is known to occur in a coupled mode model when Gaussian distribution of the uncoupled frequencies is assumed, with a width much smaller than the normal mode splitting (i.e. in the weak disorder regime) \cite{savonaprb}. From the plot it can be inferred that, in the limit $\sigma\rightarrow0$ the relation between $\braket{J}$ and $\braket{\Delta\Omega}$ becomes linear, as expected from coupled mode theory.

\begin{figure}[t]
  \begin{center}
    \includegraphics[width=0.99\textwidth]{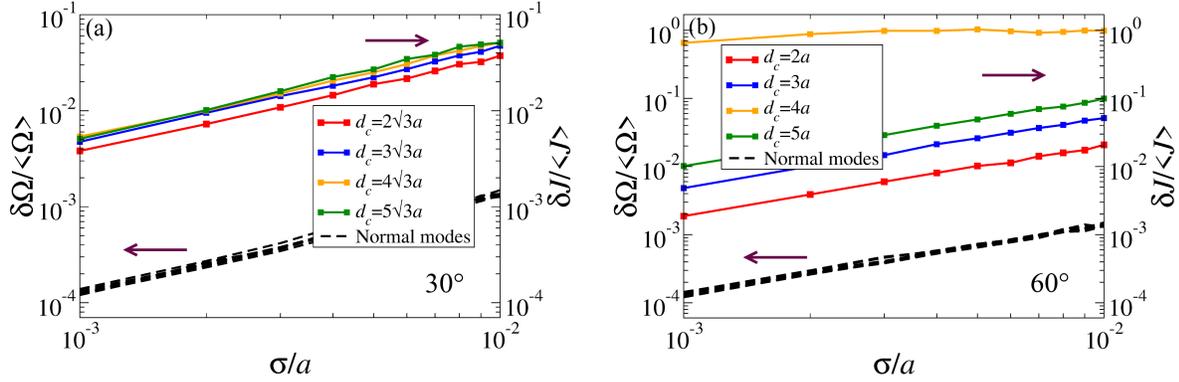}
  \end{center}
  \caption{(a) Relative fluctuations with respect to the statistical average for the normal modes, $\delta \Omega/\braket{\Omega}$ (left-axis), and coupling strength, $\delta J/\braket{J}$ (right-axis), as a function of the disorder amount $\sigma$ in the 30$^\circ$ molecule. The continuous colored curves correspond to the different cavity-cavity separations while the dashed black curves correspond to the normal modes. (b) same as (a) but for the 60$^\circ$ molecule. Each data point was computed by averaging over 500 independent statistical realizations of the disordered system.\label{fig5}}
\end{figure}

The relative fluctuations $\delta\Omega/\braket{\Omega}$ and $\delta J/\braket{J}$, are shown in Figs.~\ref{fig5}(a) and \ref{fig5}(b) for the 30$^\circ$ and 60$^\circ$ cases, respectively, as a function of the disorder amplitude. On a double logarithmic scale, we clearly identify a linear dependence of $\delta J/\braket{J}$ and $\delta\Omega/\braket{\Omega}$ on $\sigma$ (with the exception of one value of $d_c=4a$ in Fig.~\ref{fig5}(b), to be discussed below). The exponent of the resulting power law is found to be 1 within a statistical error of less than 4\%. This leads to the main conclusion of this work, namely that the relative fluctuations in the coupling strength $J$ increase linearly as a function of the disorder amplitude $\sigma$, in the same way as already known for the resonant frequencies \cite{momchil2}. We can thus write
\begin{align}
\frac{\delta J}{\braket{J}} &=\alpha\sigma, \label{Jfluc}\\
\frac{\delta\Omega}{\braket{\Omega}} &=\beta\sigma \label{Ofluc},
\end{align}
where $\alpha$ and $\beta$ depend on the specific molecule configuration. The data in Figs.~\ref{fig5}(a) and \ref{fig5}(b) clearly indicate that $\delta J/\braket{J}$ increases faster than $\delta\Omega/\braket{\Omega}$, as $\sigma$ increases. This suggests that the coupling strength is more sensitive to disorder than the normal mode frequencies. Moreover, as intuitively expected, larger values of $J$ -- usually obtained at small cavity-cavity separations -- are more robust against disorder than the smaller ones. This dependence emerges clearly when relating the normal-mode splitting in Figs.~\ref{fig1}(c) and \ref{fig1}(d), with the factor $\alpha$ obtained from Figs.~\ref{fig5}(a) and \ref{fig5}(b). It appears clearly that the smaller the normal-mode splitting, the larger $\alpha$. The extreme case with $d_c=4a$ in the 60$^\circ$ molecule corresponds to a vanishing normal-mode splitting, i.e. to a vanishing $J$. Then, the relative fluctuations in $J$ eventually become larger than the average value of $J$, and the dependence on $\sigma$ is no longer linear.

From the results displayed Fig.~\ref{fig5}, the linear dependence in Eq.~(\ref{Jfluc}) remains undoubtedly valid as long as $\delta J/\braket{J}<0.1$, which is the region where the disorder-induced fluctuations in $J$ can be considered as a small effect. This sets an upper bound to the disorder amplitude, given by $\sigma_{\rm max}=0.1/\alpha$. For typical photonic molecules as those considered here, this upper bound translates to $\sigma_{\rm max}\sim0.02a$, which is safely far from the fabrication accuracy currently achieved in Si and GaAs PhC. For designs with particularly small couplings however, such as in the 60$^\circ$ molecule at $d_c=4a$, the bound is already reached for $\sigma_{\rm max}<0.001a$. This leads to the general conclusion that, when designing a PhC structure with very small couplings, disorder may represent a major obstacle to achieving the nominal values of the coupling in fabricated structures.

\section{Conclusions} \label{conclusions}

We have studied the effects of disorder on the coupling strength of coupled photonic crystal cavities, by combining microscopic simulations of the electromagnetic modes with a coupled-mode model. We have found that the relative fluctuations of the coupling strength $J$ and normal modes frequencies $\Omega$ increase linearly with the Gaussian disorder amplitude $\sigma$. Furthermore, the disorder-induced fluctuations $\delta J/\braket{J}$ increase faster than $\delta \Omega/\braket{\Omega}$ for increasing $\sigma$, leading to the conclusion that the so called photon hopping parameter $J$ is more sensitive to random imperfections than the normal modes of the coupled cavity system. This result sets an upper bound to the amount of disorder that can be supported by coupled PhC structures, in order for the nominal coupling to be actually realized in fabricated structures. For most of the configurations studied here, we estimate this bound to lie quite above the current fabrication accuracy achieved in modern fabrication techniques for typical Si and GaAs photonic crystal slabs \cite{galli,garcia}. A more stringent bound is instead found in cases where the nominal value of $J$, as predicted by the coupled cavity design, is very small. In these cases, state-of-the-art fabrication may be insufficient to produce structures where the nominal coupling is systematically realized.

The present findings on the statistical distribution of the coupling $J$ are essentially determined by the local disorder properties in the coupling region of the PhC. The present conclusions can then be generalized to other coupled PhC structures, such as coupled cavities with different geometries or cavities coupled to waveguides. They provide general guidelines to estimate the disorder induced statistical fluctuations of the coupling strengths, when designing a coupled PhC structure.

\section*{References}
\bibliographystyle{unsrt}

\begin{thebibliography}{10}

\bibitem{yablonovitch}
Eli Yablonovitch.
\newblock {Inhibited Spontaneous Emission in Solid-State Physics and
  Electronics}.
\newblock {\em Phys. Rev. Lett.}, 58:2059, 1987.

\bibitem{john}
Sajeev John.
\newblock {Strong localization of photons in certain disordered dielectric
  superlattices}.
\newblock {\em Phys. Rev. Lett.}, 58:2486, 1987.

\bibitem{noda1}
Noda Susumu, Fujita Masayuki, and Asano Takashi.
\newblock {Spontaneous-emission control by photonic crystals and nanocavities}.
\newblock {\em Nat. Photonics}, 1:449, 2007.

\bibitem{vuckovicrev}
O'Brien~Jeremy L., Furusawa Akira, and Vu\v{c}kovi{\'c} Jelena.
\newblock {Photonic quantum technologies}.
\newblock {\em Nat. Photonics}, 3:687, 2009.

\bibitem{notomirev}
Masaya Notomi.
\newblock {Manipulating light with strongly modulated photonic crystals}.
\newblock {\em Rep. Prog. Phys.}, 73:096501, 2010.

\bibitem{gallirev}
Priolo Francesco, Gregorkiewicz Tom, Galli Matteo, and Krauss~Thomas F.
\newblock {Silicon nanostructures for photonics and photovoltaics}.
\newblock {\em Nat. Nanotechnol.}, 9:19, 2014.

\bibitem{lodahl}
Peter Lodahl, Sahand Mahmoodian, and S{\o}ren Stobbe.
\newblock {Interfacing single photons and single quantum dots with photonic
  nanostructures}.
\newblock {\em Rev. Mod. Phys.}, 87:347, 2015.

\bibitem{joannopoulosBook}
John~D. Joannopoulos, Steven~G. Johnson, Joshua~N. Winn, and Robert Meade.
\newblock {\em {Photonic Crystals: Molding the Flow of Light}}.
\newblock Princeton University Press, second edition, 2008.

\bibitem{notomi1}
Kuramochi Eiichi, Nozaki Kengo, Shinya Akihiko, Takeda Koji, Sato Tomonari,
  Matsuo Shinji, Taniyama Hideaki, Sumikura Hisashi, and Notomi Masaya.
\newblock {Large-scale integration of wavelength-addressable all-optical
  memories on a photonic crystal chip}.
\newblock {\em Nat. Photonics}, 85:474, 2014.

\bibitem{liu}
Ye~Liu, Fei Qin, Zi-Ming Meng, Fei Zhou, Qing-He Mao, and Zhi-Yuan Li.
\newblock {All-optical logic gates based on two-dimensional
  low-refractive-index nonlinear photonic crystal slabs}.
\newblock {\em Opt. Express}, 19(3):1945, 2011.

\bibitem{imamoglu}
Volz Thomas, Reinhard Andreas, Winger Martin, Badolato Antonio, Hennessy~Kevin
  J., Hu~Evelyn L., and Imamo\u{g}lu Ata\c{c}.
\newblock {Ultrafast all-optical switching by single photons}.
\newblock {\em Nat. Photonics}, 6:605, 2012.

\bibitem{notomi2}
Nozaki Kengo, Tanabe Takasumi, Shinya Akihiko, Matsuo Shinji, Sato Tomonari,
  Taniyama Hideaki, and Notomi Masaya.
\newblock {Sub-femtojoule all-optical switching using a photonic-crystal
  nanocavity}.
\newblock {\em Nat. Photonics}, 4:477, 2010.

\bibitem{noda2}
Sato Yoshiya, Tanaka Yoshinori, Upham Jeremy, Takahashi Yasushi, Asano Takashi,
  and Noda Susumu.
\newblock {Strong coupling between distant photonic nanocavities and its
  dynamic control}.
\newblock {\em Nat. Photonics}, 6:56, 2011.

\bibitem{salemink1}
Y.~Liu and H.~W.~M. Salemink.
\newblock { Photonic crystal-based all-optical on-chip sensor }.
\newblock {\em Opt. Express}, 20:19912, 2012.

\bibitem{salemink2}
Yazhao Liu and H.~W.~M. Salemink.
\newblock { All-optical on-chip sensor for high refractive index sensing }.
\newblock {\em Appl. Phys. Lett.}, 106:031116, 2015.

\bibitem{momchil1}
Momchil Minkov and Vincenzo Savona.
\newblock {Wide-band slow light in compact photonic crystal coupled-cavity
  waveguides }.
\newblock {\em Optica}, 2:631, 2015.

\bibitem{badolato}
Yiming Lai, Mohamed~Sabry Mohamed, Boshen Gao, Momchil Minkov, Robert~W. Boyd,
  Vincenzo Savona, Romuald Houdre, and Antonio Badolato.
\newblock {Ultra-wide-band slow light in photonic crystal coupled-cavity
  waveguides}.
\newblock {\em arXiv:1706.09625}.

\bibitem{momchilsr}
Momchil Minkov and Savona Vincenzo.
\newblock {Automated optimization of photonic crystal slab cavities}.
\newblock {\em Sci. Rep.}, 4:5124, 2014.

\bibitem{galli}
Y.~Lai, S.~Pirotta, G.~Urbinati, D.~Gerace, M.~Minkov, V.~Savona, A.~Badolato,
  and M.~Galli.
\newblock {Genetically designed L3 photonic crystal nanocavities with measured
  quality factor exceeding one million}.
\newblock {\em Appl. Phys. Lett.}, 104:241101, 2014.

\bibitem{momchilgan}
Noelia~Vico Trivi{\~n}o, Momchil Minkov, Giulia Urbinati, Matteo Galli,
  Jean-Fran\c{c}ois Carlin, Rapha{\"e}l Butt{\'e}, Vincenzo Savona, and Nicolas
  Grandjean.
\newblock {Gallium nitride L3 photonic crystal cavities with an average quality
  factor of 16 900 in the near infrared }.
\newblock {\em Appl. Phys. Lett.}, 105:231119, 2014.

\bibitem{ulagalandha}
Ulagalandha~Perumal Dharanipathy, Momchil Minkov, Mario Tonin, Vincenzo Savona,
  and Romuald Houdr{\'e}.
\newblock {High-Q silicon photonic crystal cavity for enhanced optical
  nonlinearities}.
\newblock {\em Appl. Phys. Lett.}, 105:101101, 2014.

\bibitem{momchilwv}
Momchil Minkov and Vincenzo Savona.
\newblock {Wide-band slow light in compact photonic crystal coupled-cavity
  waveguides}.
\newblock {\em Optica}, 2:631, 2015.

\bibitem{momchill43}
Momchil Minkov, Vincenzo Savona, and Dario Gerace.
\newblock {Photonic crystal slab cavity simultaneously optimized for ultra-high
  Q/V and vertical radiation coupling}.
\newblock {\em Appl. Phys. Lett.}, 111:131104, 2017.

\bibitem{mohamed}
Mohamed~Sabry Mohamed, Angelica Simbula, Jean-Fran\c{c}ois Carlin, Momchil
  Minkov, Dario Gerace, Vincenzo Savona, Nicolas Grandjean, Matteo Galli, and
  Romuald Houdr{\'e}.
\newblock {Efficient continuous-wave nonlinear frequency conversion in high-Q
  gallium nitride photonic crystal cavities on silicon}.
\newblock {\em APL Photonics}, 2:031301, 2017.

\bibitem{ramunno}
L.~Ramunno and S.~Hughes.
\newblock {Disorder-induced resonance shifts in high-index-contrast photonic
  crystal nanocavities}.
\newblock {\em Phys. Rev. B}, 79:161303(R), 2009.

\bibitem{momchil2}
Momchil Minkov, Ulagalandha~Perumal Dharanipathy, Romuald Houdr{\'e}, and
  Vincenzo Savona.
\newblock {Statistics of the disorder-induced losses of high-Q photonic crystal
  cavities }.
\newblock {\em Opt. Express}, 21:28233, 2013.

\bibitem{vasco1}
Juan~Pablo Vasco and Stephen Hughes.
\newblock {Anderson Localization in Disordered LN Photonic Crystal Slab
  Cavities}.
\newblock {\em ACS Photonics}, 5:1262, 2018.

\bibitem{dario}
Dario Gerace and Lucio~Claudio Andreani.
\newblock {Disorder-induced losses in photonic crystal waveguides with line
  defects }.
\newblock {\em Opt. Lett.}, 29:1897, 2004.

\bibitem{hughes}
S.~Hughes, L.~Ramunno, Jeff~F. Young, and J.~E. Sipe.
\newblock {Extrinsic Optical Scattering Loss in Photonic Crystal Waveguides:
  Role of Fabrication Disorder and Photon Group Velocity}.
\newblock {\em Phys. Rev. Lett.}, 94:033903, 2005.

\bibitem{momchil3}
Momchil Minkov and Vincenzo Savona.
\newblock {Effect of hole-shape irregularities on photonic crystal waveguides
  }.
\newblock {\em Opt. Lett.}, 37:3108, 2012.

\bibitem{vasco2}
J.~P. Vasco and S.~Hughes.
\newblock {Statistics of Anderson-localized modes in disordered photonic
  crystal slab waveguides}.
\newblock {\em Phys. Rev. B}, 95:224202, 2017.

\bibitem{dignam}
D.~P. Fussell, S.~Hughes, and M.~M. Dignam.
\newblock {Influence of fabrication disorder on the optical properties of
  coupled-cavity photonic crystal waveguides}.
\newblock {\em Phys. Rev. B}, 78:144201, 2008.

\bibitem{savona1}
Vincenzo Savona.
\newblock {Electromagnetic modes of a disordered photonic crystal}.
\newblock {\em Phys. Rev. B}, 83:085301, 2011.

\bibitem{vuckovic}
Arka Majumdar, Armand Rundquist, Michal Bajcsy, Vaishno~D. Dasika, Seth~R.
  Bank, and Jelena Vu\v{c}kovi{\'c}.
\newblock {Design and analysis of photonic crystal coupled cavity arrays for
  quantum simulation}.
\newblock {\em Phys. Rev. B}, 86:195312, 2012.

\bibitem{hartmann1}
M~Hartmann, F~Brand{\~a}o, and M~Plenio.
\newblock {Quantum many‐body phenomena in coupled cavity arrays}.
\newblock {\em Laser Photon. Rev.}, 2:527, 2008.

\bibitem{savona2}
T.~C.~H. Liew and V.~Savona.
\newblock {Quantum entanglement in nanocavity arrays}.
\newblock {\em Phys. Rev. A}, 85:050301(R), 2012.

\bibitem{cristiano}
Iacopo Carusotto and Cristiano Ciuti.
\newblock {Quantum fluids of light}.
\newblock {\em Rev. Mod. Phys.}, 85:299, 2013.

\bibitem{hugo}
Flayac Hugo, Gerace Dario, and Savona Vincenzo.
\newblock {An all-silicon single-photon source by unconventional photon
  blockade}.
\newblock {\em Sci. Rep.}, 5:11223, 2015.

\bibitem{vasco3}
J.~P. Vasco, D.~Gerace, P.~S.~S. Guimar{\~a}es, and M.~F. Santos.
\newblock {Steady-state entanglement between distant quantum dots in photonic
  crystal dimers}.
\newblock {\em Phys. Rev. B}, 94:165302, 2016.

\bibitem{angelakis}
Changsuk Noh and Dimitris~G Angelakis.
\newblock {Quantum simulations and many-body physics with light}.
\newblock {\em Rep. Prog. Phys.}, 80(1):016401, 2017.

\bibitem{gme}
Lucio~Claudio Andreani and Dario Gerace.
\newblock {Photonic-crystal slabs with a triangular lattice of triangular holes
  investigated using a guided-mode expansion method}.
\newblock {\em Phys. Rev. B}, 73:235114, 2006.

\bibitem{vasco4}
J.P. Vasco, P.~S.~S. Guimar{\~a}es, and D.~Gerace.
\newblock {Long-distance radiative coupling between quantum dots in photonic
  crystal dimers}.
\newblock {\em Phys. Rev. B}, 90:155436, 2014.

\bibitem{chalcraft}
A.~R.~A. Chalcraft, S.~Lam, B.~D. Jones, D.~Szymanski, R.~Oulton, A.~C.~T.
  Thijssen, M.~S. Skolnick, D.~M. Whittaker, T.~F. Krauss, and A.~M. Fox.
\newblock {Mode structure of coupled L3 photonic crystal cavities}.
\newblock {\em Opt. Express}, 19:5670, 2011.

\bibitem{nishan}
Nishan Mann, Alisa Javadi, P.~D. Garc{\'i}a, Peter Lodahl, and Stephen Hughes.
\newblock {Theory and experiments of disorder-induced resonance shifts and
  mode-edge broadening in deliberately disordered photonic crystal waveguides}.
\newblock {\em Phys. Rev. A}, 92:023849, 2015.

\bibitem{thomas}
N.~Le Thomas, Z.~Diao, H.~Zhang, and R.~Houdr{\'e}.
\newblock {Statistical analysis of subnanometer residual disorder in photonic
  crystal waveguides: Correlation between slow light properties and structural
  properties }.
\newblock {\em J. Vac. Sci. Technol. B}, 29:051601, 2011.

\bibitem{garcia}
P.~D. Garc{\'i}a, A.~Javadi, H.~Thyrrestrup, and P.~Lodahl.
\newblock {Quantifying the intrinsic amount of fabrication disorder in
  photonic-crystal waveguides from optical far-field intensity measurements}.
\newblock {\em Appl. Phys. Lett.}, 102:031101, 2013.

\bibitem{savonaprb}
V.~Savona and C.~Weisbuch.
\newblock {Theory of time-resolved light emission from polaritons in a
  semiconductor microcavity under resonant excitation}.
\newblock {\em Phys. Rev. B}, 54:10835, 1996.

\end{thebibliography}

\end{document}